\documentclass[%
aip,
amsmath,amssymb,
reprint,%
]{revtex4-1}

\usepackage{graphicx}
\usepackage{dcolumn}
\usepackage{bm}
\usepackage[utf8]{inputenc}
\usepackage[T1]{fontenc}
\usepackage{mathptmx}
\usepackage{etoolbox}
\usepackage[colorlinks=true,linkcolor=blue]{hyperref}%
\makeatletter
\def\@email#1#2{%
	\endgroup
	\patchcmd{\titleblock@produce}
	{\frontmatter@RRAPformat}
	{\frontmatter@RRAPformat{\produce@RRAP{*#1\href{mailto:#2}{#2}}}\frontmatter@RRAPformat}
	{}{}
}%
\makeatother
\begin{document}
	
	\preprint{AIP/123-QED}
	
	\title{A novel machine learning enabled hybrid optimization framework for efficient and transferable coarse-graining of a model polymer}
	\author{Zakiya Shireen}\thanks{Contributed equally in carrying out the work and writing the manuscript.} 
	\author{Hansani Weeratunge}\thanks{Contributed equally in carrying out the work and writing the manuscript.} 
	\email{ellie.hajizadeh@unimelb.edu.au.}
	\affiliation{Department of Mechanical Engineering, Faculty of Engineering and Information Technology, The University of Melbourne, Melbourne, Australia}
	\author{Adrian Menzel}
	\author{Andrew W Phillips}
	\affiliation{Maritime Division, Defence Science and Technology Group, Melbourne, Australia}
	\author{Ronald G Larson}
	\affiliation{Department of Chemical Engineering, University of Michigan, Ann Arbor, Michigan, USA}
	\author{Kate Smith-Miles}
	\affiliation{School of Mathematics and Statistics, The University of Melbourne, Melbourne, Australia}
	\author{Elnaz Hajizadeh}
	\affiliation{Department of Mechanical Engineering, Faculty of Engineering and Information Technology, The University of Melbourne, Melbourne, Australia}

	\date{\today}

	
	\begin{abstract}
		This work presents a novel framework governing the development of an efficient, accurate, and transferable coarse-grained (CG) model of a polyether material. The proposed framework combines the two fundamentally different classical optimization approaches for the development of coarse-grained model parameters; namely bottom-up and top-down approaches. This is achieved through integrating the optimization algorithms into a machine learning (ML) model, trained using molecular dynamics (MD) simulation data. In the bottom-up approach, bonded interactions of the CG model are optimized using deep neural networks (DNN), where atomistic bonded distributions are matched. The atomistic distributions emulate the local chain structure. In the top-down approach, optimization of nonbonded potentials is accomplished by reproducing the temperature-dependent experimental density. We demonstrate that CG model parameters achieved through our machine-learning enabled hybrid optimization framework fulfills the thermodynamic consistency and transferability issues associated with the classical approaches to coarse-graining model polymers. We demonstrate the efficiency,  accuracy, and transferability of the developed CG model, using our novel framework through accurate predictions of chain size as well as chain dynamics, including the limiting behavior of the glass transition temperature, diffusion, and stress relaxation spectrum, where none were included in the potential parameterization process. The accuracy of the predicted properties are evaluated in the context of molecular theories and available experimental data.
		
	\end{abstract}
	
	\maketitle
	
	\section{\label{sec:level1}Introduction}
	
	Molecular dynamics (MD) simulation techniques provide a powerful route to establish the structure-property relationships in materials through solving the coupled equations of motions of interacting atoms and molecules in a material system \cite {behbahani2021dynamics,grest2020resolving,lempesis2016atomistic}. Despite the great success of this computational technique, modeling of the macromolecules, such as proteins and polymers across multiple lengths and time scales is restrained by computational limitations. To overcome the challenges of modeling a macromolecular system such as polymers for a longer time, coarse-grained models (CG) are required. One of the key features of coarse-graining is the ability to probe polymer behaviour over large spatiotemporal scales, which is otherwise difficult to achieve in high-resolution models.
	The central problem in coarse-graining polymeric materials is to retain the chain attributes of interest while constructing the adequate representation of pseudo atom or CG bead. One of the defining characteristics of polymer materials is the long time dynamical response that occurs at multiple length scales. Coarse-graining allows to combine groups of atoms into fewer interaction sites, thus, reducing the degrees of freedom in the system. In addition, the accurate representation requires encoding of finer details, i.e., the chemical specificity should be incorporated into the CG bead descriptor. The CG model can be made to mimic the atomistic structural features by optimizing the interaction parameters by matching the atomistic distributions, the so-called bottom-up approach. Contrary, in the top-down strategy the chemical specificity is subsumed into the descriptor by matching the macroscopic properties such as density ($\rho$), glass transition temperature ($T_g$), and elastic modulus ($E$) etc \cite{hsu2015thermomechanically}.
	
	The classical methods such as iterative Boltzmann inversion (IBI) \cite{agrawal2014simultaneous,liu2019coarse, bayramoglu2012coarse,ohkuma2020composition}, inverse Monte Carlo \cite{korolev2014coarse,lyubartsev2015systematic}, and relative entropy \cite{foley2015impact,shell2016coarse} typically aim to map the structural distribution of atomistic model. However, these structure-based conventional methods have limitations in capturing the correct thermodynamic properties and free energy landscape. While the force-matching methods such as many-body potential of mean force (PMF) capture the accurate dynamics but it represents the atomistic structural features inaccurately \cite{dunn2016van,noid2013perspective,brini2013systematic}. Optimization of nonbonded interactions in a bottom-up manner i.e. by optimizing the radial distribution functions (RDF), usually results in larger deviations of the thermal expansion coefficient ($\alpha$). Hence, a purely bottom-up optimized CG model does not demonstrate thermomechanical consistency, therefore, temperature-dependent density transferability can not be guaranteed a priori \cite{louis2002beware,fritz2009coarse,agrawal2014simultaneous}. The hybrid approach, i.e., combining the bottom-up and top-down strategies in classical CG models has been developed for a couple of materials systems \cite{huang2018transferrable,hsu2015thermomechanically}. Coarse-graining of polystyrene was studied by adopting the top-down strategy, where nonbonded potential parameters were optimized by increasing the pairwise potential energy well depth. The increased well depth reintroduces a rough additive potential energy landscape contrary to the usual smooth potential energy landscape of CG models \cite{hsu2015thermomechanically}. However, using liquid perturbation theory, it has been demonstrated that for common polyethers the nonbonded interactions in the coarse-grained models are temperature dependent \cite{huang2018transferrable}.
	
	Although the conventional CG models are straightforward and allow much larger systems with reasonable computation costs \cite{hajizadeh2014shear, hajizadeh2015molecular}, they still suffer from some important issues in terms of thermomechanical consistency, representability and transferability \cite{dunn2016van,li2013challenges}. Therefore, although the classical hybrid approach has improved the accuracy of the CG model predictions, it throws challenges, such as careful fine tuning of the nonbonded $LJ$ potential parameters (well depth $\epsilon$) and variations in auxiliary $LJ$ potential terms. The fine tuning of pairwise interaction parameters can be very demanding and time consuming for complex block copolymers and biological macromolecules \cite{voth2008coarse,kmiecik2016coarse,singh2019recent} due to multiple pairwise interaction sites and inherent iterative process of optimization. 
	
	To address the questions pertaining to generic coarse-graining mechanisms of complex molecules and accelerate the process, data-driven and machine learned (ML) potentials is emerging as an alternative approach. The machine learned CG potentials have been proposed to present free-energy landscape of all-atoms (AA) molecular model with enhanced efficiency and accuracy \cite{ye2021machine}. Thus, the ML enabled CG potentials are becoming a tool to bridge the gap between accurate and computationally expensive all-atomistic method, and approximate but computationally affordable CG methods \cite{behler2011atom, ramprasad2017machine, chan2019machine, moradzadeh2019transfer, nguyen2021integration, duan2019machine, mcdonagh2019utilizing, hajizadeh2014nonequilibrium}. It is demonstrated elsewhere that the machine-learned CG potentials can predict the physical properties of complex molecular systems with \textit{ab-initio} accuracy \cite{ruza2020temperature,wang2019machine,zhang2018deepcg}. To reduce the computational cost, machine learning and optimization techniques such as particle swarm optimization (PSO), genetic algorithm (GA), and simplex method \cite{chan2019machine, bejagam2018machine,reith2001mapping} are integrated to enhance the efficiency of these methods. The emerging data-driven technique \cite {kuenneth2021copolymer,chen2021polymer} is promising and is credible to achieve accuracy with enhanced efficiency compared to the conventional approaches \cite{prathumrat2022shape}.	
	
	The potential of statistical learning techniques such as neural networks lies in capturing the hidden representations of complex data. They are proven to be capable of formulating the complex potentials using all-atoms molecular dynamics simulation data as reference or ground truth \cite{wang2019coarse}. Similar to classical CG potentials, the machine-learned CG (MLCG) methods are also categorised as: bottom-up MLCG methods \cite{moradzadeh2019transfer,chan2019machine,bejagam2018machine} and top-down MLCG methods \cite{jeong2022extended,karuth2021predicting,scherer2020kernel,webb2018graph}. Generally, in the bottom-up MLCG methods, a CG potential that is related to representations of CG beads is introduced through multi layer neural networks \cite{wang2019coarse}. Then the neural network is trained and optimized by utilizing the reference system i.e. data set from the all-atoms molecular simulations. In the top-down MLCG method, an empirical relation of the interactions between CG beads (eg. LJ potential for a simple fluid) is given as a priori \cite{jeong2022extended,chan2019machine,moradzadeh2019transfer}. The interaction parameters in the empirical relation are then optimized by a statistical learning technique to match features such as radial distribution function (RDF) and temperature-dependent volumetric behaviour with corresponding simulations/experimental results.
	
	In this work, we propose a novel framework by combining bottom-up and top-down MLCG methods that enable thermodynamic consistency and transferability, while maintaining the structural representability. 
	Further, we demonstrate the application of the proposed method by coarse-graining poly(tetramethylene oxide) (PTMO), which is one of the major components of elastomeric polyurethanes. The coarse-grained poly(tetramethylene oxide) (PTMO) model represents the soft segment of the polyurethane (PU) chain. The framework involves the integration of MLCG methods with genetic algorithm (GA) and molecular dynamics (MD) simulations. In the developed protocol the bottom-up MLCG approach is used to establish the bonded interactions by training deep neural networks (DNN) and top-down MLCG is used by incorporating GA optimization scheme to obtain nonbonded interaction parameters by matching temperature-dependent density. One of the key highlights of the proposed framework is that the MD simulations are replaced by a DNN that is trained to predict the temperature-dependent density to further accelerate the optimization component. Therefore, we avoid the need for a large number of simulations, where data are generally not reused. Thus, DNNs are used as surrogate models, which addresses the re-usability issue of the data \cite{weeratunge2022machine}. Our focus is on the working mechanism of the framework by demonstrating the ability of the CG model to make accurate predictions, and the validation of the trained model.
	
	\section{\label{sec:level2} Characterization of the MLCG framework}
	\begin{figure*}[!ht]
		\centering
		\includegraphics[width=0.70\textwidth]{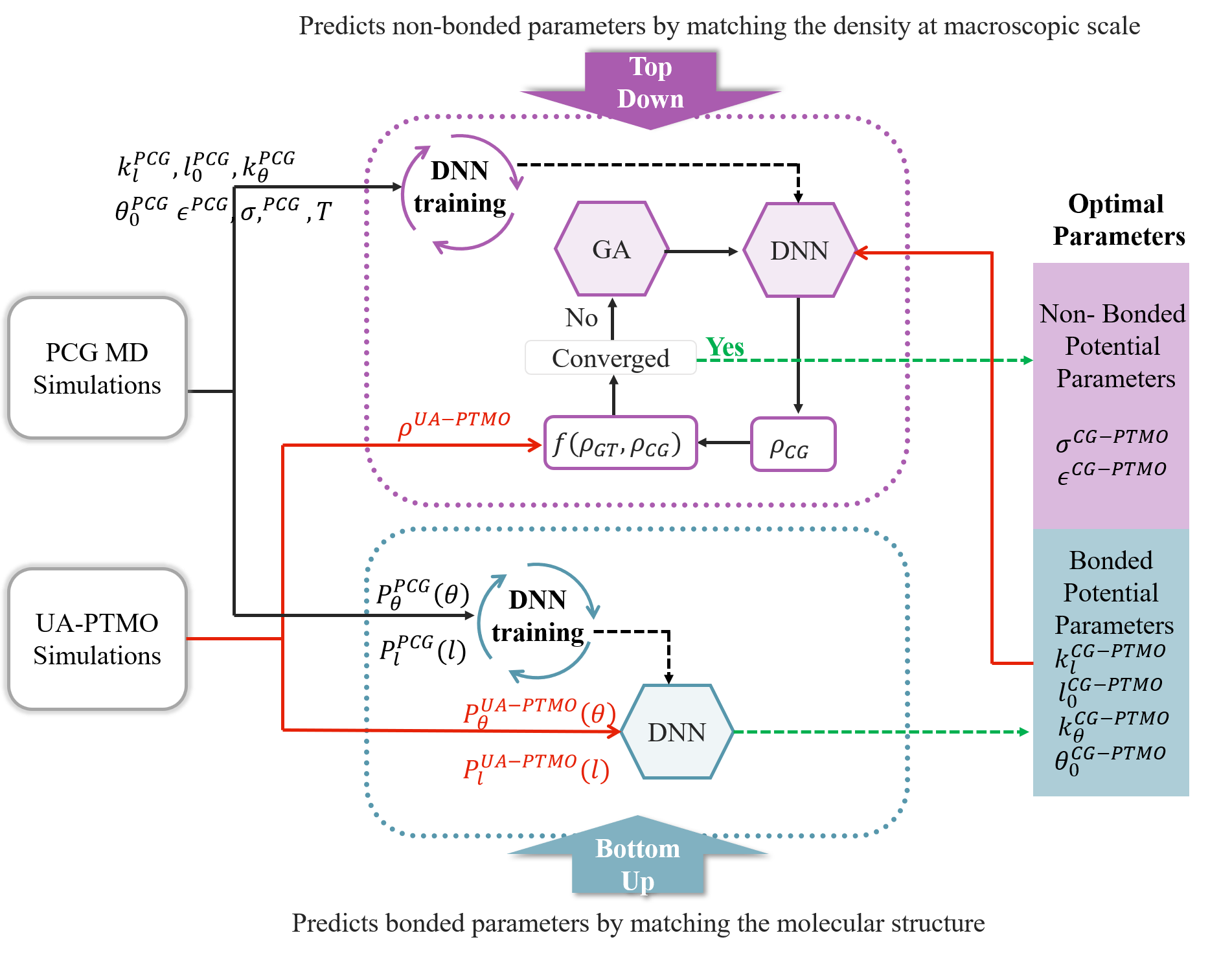}
		\caption{Schematic representation of the framework, where a hybrid optimization approach has been used to develop interaction parameters for coarse-graining polymer molecules using machine learning technique. The framework includes the bottom-up optimization approach (detail given in \ref{sec:level2.1}) for predicting bonded interaction parameters ($k_l^{CG-PTMO}, l_0^{CG-PTMO}, k_\theta^{CG-PTMO}, \theta_0^{CG-PTMO}$) by matching bond ($P_l^{UA-PTMO}$),  and angle distributions ($P_\theta^{UA-PTMO}$) from united-atom model of PTMO as ground truth. In bonded optimization process, the DNN was trained using bond ($P_l^{PCG}(l)$) and angle ($P_\theta^{PCG}(\theta)$) distribution data from prototype-coarse-grained (PCG) simulations. The top-down approach (\ref{sec:level2.2}) was incorporated with genetic algorithm and deep neural network for predicting the nonbonded interaction parameters ($\sigma^{CG-PTMO}, \epsilon^{CG-PTMO}$) of CG model. The DNN was trained using parameters ($k_l^{PCG}, l_0^{PCG}, k_\theta^{PCG}, \theta_0^{PCG}, \epsilon^{PCG}, \sigma^{PCG}$, and $T$) acquired from PCG simulations and was integrated into GA along with parameters obtained in bottom-up step to predict the density. Consecutively, nonbonded interaction parameters ($\epsilon^{CG-PTMO}, \sigma^{CG-PTMO}$)  were optimized by matching temperature-dependent experimental density and density data from united-atom ($\rho^{UA-PTMO}$) simulations.}
		\label{Figure:1}
	\end{figure*}

	We have established a novel framework through integrating machine learning, optimization, and MD simulations as illustrated in Figure \ref{Figure:1} to develop a temperature-transferable coarse-grained (CG) model that accurately represents poly(tetramethylene oxide).
	In the present study, target averaged distributions (bonded and nonbonded) are generated from the united-atoms (UA) molecular dynamics simulations. A single bead CG model is utilized, where each bead represents a repeating unit i.e. $-(CH_2CH_2OCH_2CH_2)-$ of PTMO along the polymer chain. The resulting CG bead-chains can therefore be denoted by chains of $n$ beads, which corresponds to the number of repeat units in the united-atom resolution of PTMO (see Appendix \ref{UAS} for details). Each CG bead is assigned a mass equal to the sum of the atomistic masses of its constituent elements, which is $m = 72$ g/mol. 
	
	In order to capture the structure-property relation of CG-PTMO, we use the "Prototype" coarse-grained (PCG) systems (Prototype is called PCG throughout this report) for data acquisition. Thousands of independent MD simulations were run for PCG, to obtain the data for training and testing purposes. These simulations were run using the potential energy ($U_{total}$) described by Eqs. \ref{eq1}. To explore the parameter space, the bonded and nonbonded potential parameters were randomly sampled over a specific range considering the physics of the system and the UA-PTMO parameter values.
	\begin{equation}
	U_{total} = U_{stretch} + U_{bend} + U_{nonb}
	\label{eq1}
	\end{equation}

	The total potential energy of the PCG, $U_{total}$ is given by three contributions for which the functional forms are given in Table \ref{Table:2}. $U_{stretch}$ is the energy associated with stretching the bonds with equilibrium bond length $l_0$ between adjacent beads along the polymer chain, $U_{bend}$ is the energy associated with bending i.e. the angles subtended by consecutive bonds with equilibrium angle $\theta_0$, and $U_{nonb}$ is the energy of nonbonded interactions between beads of inter chains or beads separated by more than three bonds on the same chain. Thus, the target (CG-PTMO) model demands optimization of a total of $6$ potential parameters ($k_l, l_{0}, \theta, k_{\theta}, \sigma$ and $\epsilon$) to fully describe the energetics and conformational dynamics of the system. The potential parameters of the target CG system are determined in two steps. In the first step, the bottom-up approach is used to establish the bonded interaction parameters. In the second step, the nonbonded parameters are determined from the top-down approach involving the temperature-dependent volumetric behavior of the system.
		
	\subsection{\label{sec:level2.1}Bottom-Up MLCG: Bonded Interaction Optimization}
	\begin{figure*}[!ht]
		\centering
		\includegraphics[width=0.85\textwidth]{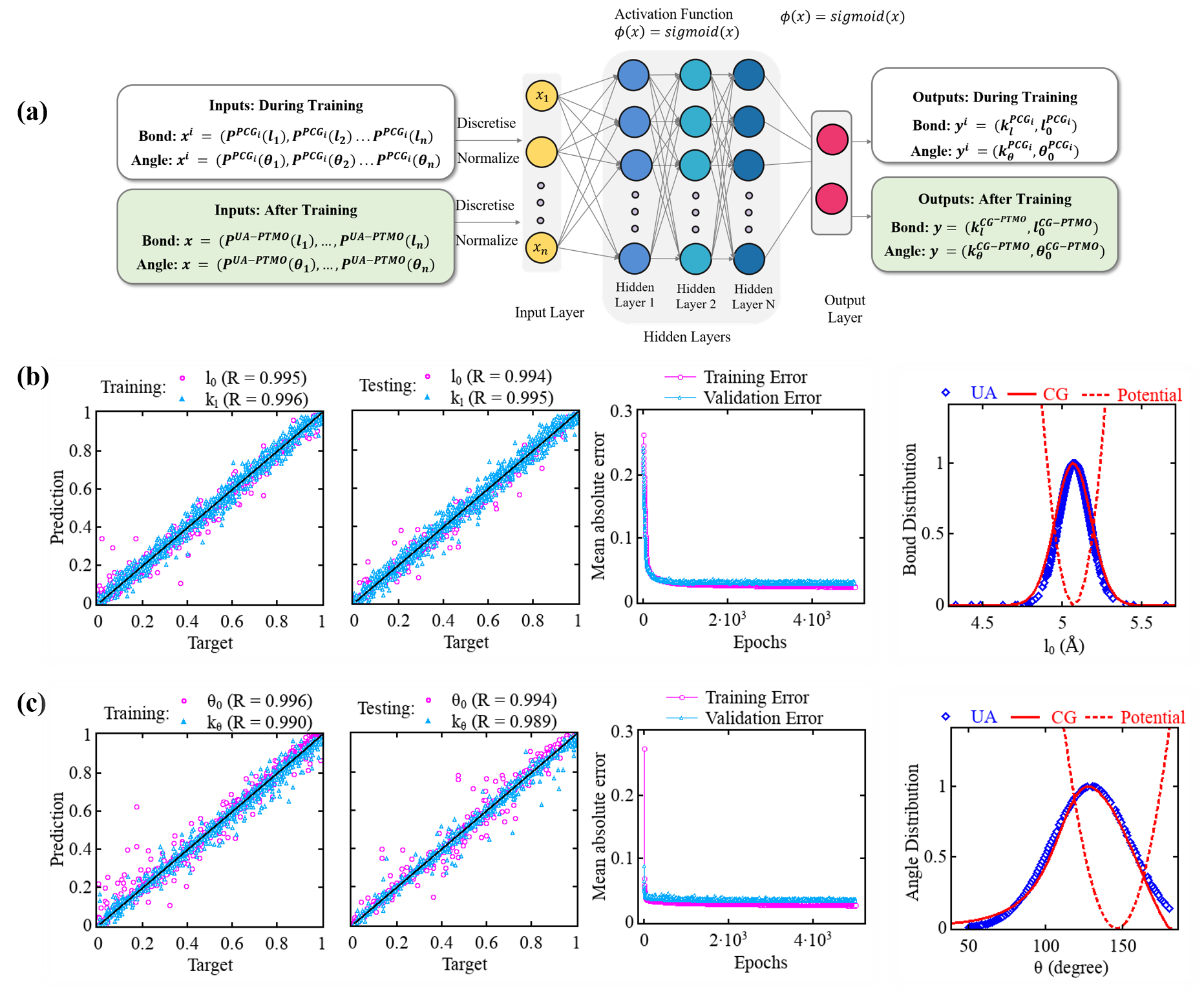}
		\caption{(a) The architecture of the DNN used in obtaining the bonded potential parameters. (b) The scatter plots represent the predictions of the DNN and the target of training and testing data for bond length ($l_0$) and stretching coefficient ($k_l$) along with the performance analysis of the neural network during training, and corresponding potential and bond distribution from MLCG predictions. (c) The target and the predictions from the DNN for angle ($\theta_0$) and bending coefficient ($k_{\theta}$) along with the performance of the DNN during training and corresponding potential and angle distribution from MLCG and ground truth.}
		\label{Figure:2}
	\end{figure*}
	Multi-layer feed forward DNNs were developed to learn the relationship between the bonded interaction parameters and local chain structures. The number of layers, nodes, and hyperparameters were tuned using learning curves. The expression of the complex relationship between the underlying potential and structural features is given in Eq. \ref{eq5} - Eq. \ref{eq6}.
	
	\begin{equation}
	(k_l,l_0 )=f(P_b (l))
	\label{eq5}
	\end{equation}
	
	\begin{equation}
	(k_\theta,\theta_0 )=f(P_\theta (\theta))
	\label{eq6}
	\end{equation}
	
	The extensive training data set of energies and structural properties of PTMO are taken from $5000$ independent PCG systems with random sampling over a broad range of bond and angle potential parameters. The data are normalized and randomly split into two sets. 70\% of the data was used to train the model and, 20\% of this training data was randomly selected for validation. The remaining 30\% of data is used for testing. Therefore, the generality of the network and its performance are evaluated for unseen data. Two distinct DNNs with densely connected layers were trained to predict the bonded potential parameters of the CG-PTMO from the bond and angle distributions as shown in Figures \ref{Figure:2} (b) and (c). The accuracy of the network is assessed using the mean absolute error (MAE) that calculates the deviation between the predicted and target values.  The network is optimized using the Adam optimizer, which is a stochastic gradient descent algorithm based on adaptive estimates of lower-order moments \cite{kingma2014adam}. 
	
	By adopting the bottom-up machine-learned coarse-graining (MLCG) approach, the data acquired from PCG simulations are fed into the network. The architecture of the developed DNN is shown in \ref{Figure:2} (a). The input vector $x^i$ is the distribution of the $i^{th}$ PCG system, which are discretized between its minimum and maximum range, i.e., $x^i_l = P(l_{min}<l<l_{max})$, and $x^i_\theta = P(\theta_{min}<\theta<\theta_{max})$. The DNN for the bond distribution has four hidden layers with $80$ nodes for each layer, with a sigmoid activation function at each layer. The output layer consists of two nodes for $k_l$, and $l_0$. The DNN developed for the angle potential consists of three hidden layers, with $60$ nodes and sigmoid activation function at each layer. The two nodes in the output layer correspond to $k_\theta$ and $\theta_0$.
	
	\begin{table*}[!ht]
		\caption{\label{Table:1} Performance of the Deep Neural Networks (DNNs).}
		\begin{ruledtabular}
			\begin{tabular}{ccccc}
				DNN& MAE (Training)&MAE (Validation)&MAE (Testing)& Training time (s)\\
				Bonded Potential\footnotemark[1] & 0.025& 0.029&  0.028 & 1539\\
				Angle Potential\footnotemark[1] & 0.026 &0.034  &0.03 & 963 \\
				Density &0.029& 0.033&0.036 & 1952 \\
			\end{tabular}
		\end{ruledtabular}	
		\footnotetext[1]{The DNNs of bonded and angle potentials present the aggregated mean absolute errors}
	\end{table*}
	
	\begin{figure*}[!ht]
		\centering
		\includegraphics[width=1.0\textwidth]{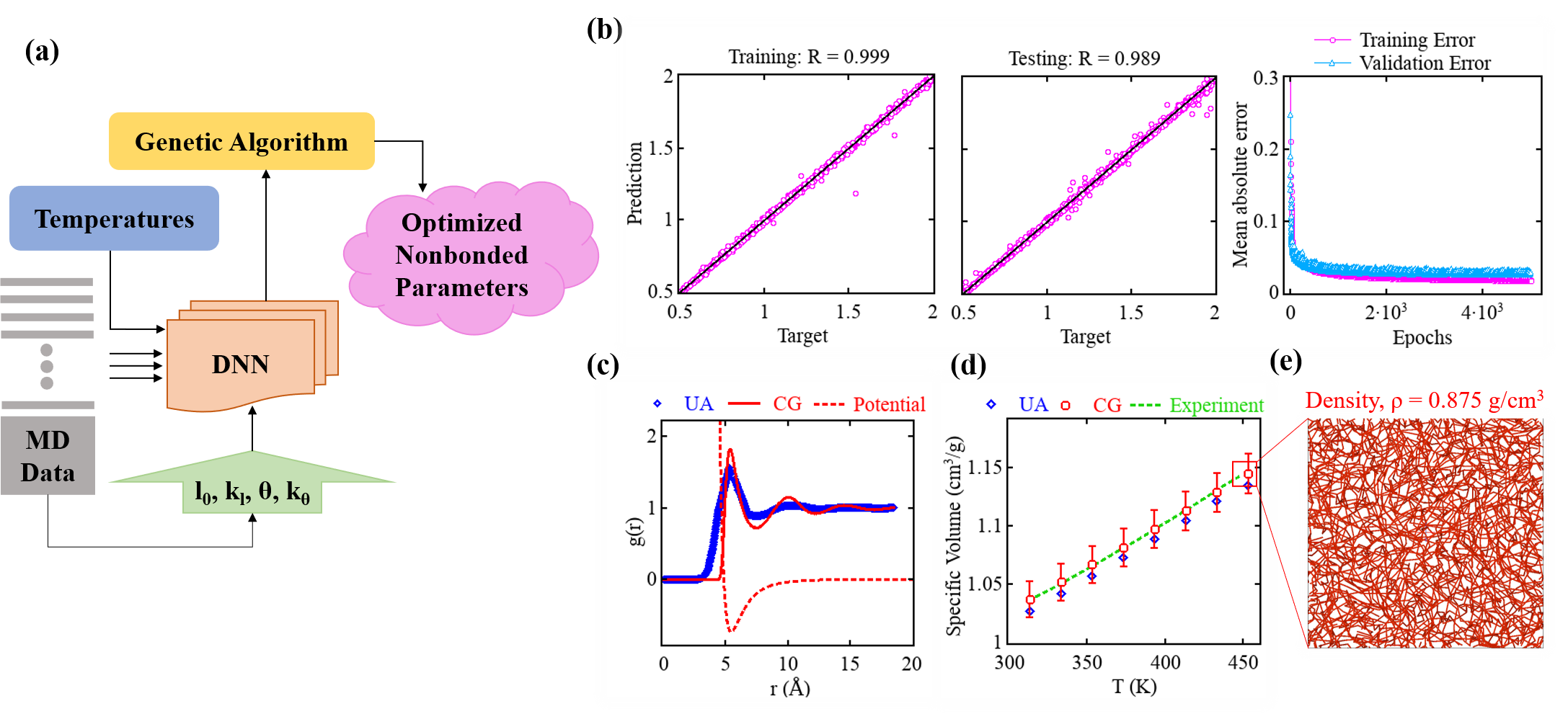}
		\caption{(a) The schematic of top-down optimization protocol (b) The scatter plots represent the predictions from the DNN and the target for training and testing data for density ($\rho_{CG}$), along with the performance of the DNN during training. (c) From the predicted parameters ($\sigma$ and $\epsilon$) the resulting radial distribution function ($g(r)$) of CG system is accompanying the corresponding potential and RDF of UA system. (d) The specific volume of MLCG system is shown along with ground truth i.e experiment (dashed line) and UA PTMO simulations. (e) Snapshot of the CG PTMO system with equilibrated density, $\rho = 0.875$ $g/cm^3$ at $T = 453$ K.}
		\label{Figure:3}
	\end{figure*}	

	Once the training of the neural networks is accomplished, the bond and angle distributions from the united-atom MD simulations, i.e., "the ground truth" are fed into the networks to predict the corresponding bonded potential parameters. This way, the DNN provides the bonded interaction parameters of the CG-PTMO model that represents the same bond distributions as the UA-PTMO model.
	
	As shown in Figure \ref{Figure:2} (b) and (c) The bond and angle distributions of the CG system exhibit a good match with reference values from the UA system, suggesting that the developed stretching and bending potentials, using the deep neural networks successfully represent the PTMO chains. The dashed lines represent the developed potentials obtained via equations listed in Table \ref{Table:2} for bonded contributions. In Table \ref{Table:2}, we have also listed the readily available potential parameters for PTMO from the literature. Please note parameters taken from literature for PTMO are for the soft segment of the polyurea \cite{agrawal2014simultaneous}. The CG predictions of the current study are not directly comparable with CG model by using these potentials due to the unavailability of the nonbonded interaction parameters.
	
	\subsection{\label{sec:level2.2}Top-Down MLCG: Nonbonded Interaction Optimization}
	
	In the conventional CG MD simulations, developing transferable force field parameters have been a longstanding challenge particularly for complex macromolecules. The traditional approaches involve the optimization of the radial distribution function, which often requires an additional correction term to include pressure fluctuations for accurate prediction of the thermodynamical behavior. In our framework, to predict the accurate thermodynamics across broad temperature ranges, the nonbonded interaction parameters ($\sigma$ and $\epsilon$) are determined by supplementing the density of poly(tetramethylene oxide) from the experiments, and united-atom simulation data along with the bonded potential parameters obtained in the bottom up approach. 
	\begin{table*}
		\caption{\label{Table:2} Summary of predicted interaction parameters from Machine Learned Coarse-Grained (MLCG) model of PTMO.}
		\begin{ruledtabular}
			\begin{tabular}{ccccc}
				&&&{Present Work}&{Literature} \\
				&Functional Form&CG Parameters&(MLCG)&(IBI) \\\hline
				Bonded&$U_{stretch}(r) = k_{l}\cdot(l-l_{0})^2$&$l_0$ (\AA)&$5.07$ &$4.98$ \\
				(Bottom-up)&&$k_l$ $(kcal/mol{\AA}^2)$&$35.95$&$3.03$ \\
				&$U_{bend}(\theta) = k_{\theta}\cdot(\theta-\theta_{0})^2$&$\theta(deg)$&$146.11$&$169.5$ \\
				&&$k_\theta$ $(kcal/mol{rad^2})$&$1.21$&$0.64$ \\
				\hline
				Nonbonded&$U_{nonb}(r_{ij}) = 4\epsilon_{ij} \left[\left({\frac{\sigma_{ij}}{r_{ij}}}\right)^{12} - \left({\frac{\sigma_{ij}}{r_{ij}}} \right)^6\right] $&$\sigma$&$4.82$&$-$ \\
				(Top-down)&&$\epsilon$& 0.73&$-$ \\
			\end{tabular}
		\end{ruledtabular}	
	\end{table*}
		
	A genetic algorithm (GA) based optimization scheme is introduced to map the density of the CG model at different thermodynamic states. The objective function of the algorithm is given as Eq. \ref{eq:ob}, where $T_i$ is the $i$th temperature state, $\rho_{CG}$ and $\rho_{GT}$ are the densities of the CG systems and ground truth (GT), respectively. The objective function evaluates the error between the supplemented density data (the ground truth) and predicted density from the CG model.
		
	\begin{equation}
	\text{Objective Function:  min} \sum_i\Big(\frac{\rho_{CG}(T_i)}{\rho_{GT}(T_i)} -1\Big)^2 
	\label{eq:ob}
	\end{equation}
	
	Since the GA is a meta-heuristic search algorithm based on the natural selection of a population with the process of adaptation for survival, it is a robust and efficient technique to explore complex nonlinear solution space. Compared to other algorithms, such as swarm intelligence-based optimization techniques, GA is less likely to have a premature convergence to a local optimal solution \cite{katoch2021review}.
	
	Due to the iterative process of optimization algorithms, generally, several hundred thousands of MD simulations are required to capture the reasonable range in the data. Therefore, approximations or surrogate models with less computational costs can be constructed to replace the MD simulations.
	In our framework, we have constructed a DNN as a surrogate model to establish the relationship between the CG potential parameters and the temperature-dependent density of the CG model. As a result, the need for further data acquisition for density from computationally expensive MD simulations has been avoided. 
	
	To train the deep neural network, bonded potential parameters of the PCG-MD simulations are fed along with temperatures to learn the multiplex relation between density and potential parameters. The multiplex expression is given in Eq. \ref{eq:7}, where $\rho$ is the density, and $\sigma, \epsilon, k_l, l_0, k_\theta, \theta_0 $ are the potential parameters.
	
	\begin{equation}
	\rho(T)=f(\sigma,\epsilon,k_l,l_0,k_\theta,\theta_0 )
	\label{eq:7}
	\end{equation}
	
	This DNN consists of three hidden layers, each with $40$ nodes and rectified linear unit (Relu) activation function. The trained DNN is then integrated into the GA optimization algorithm to expedite the exploration of the optimal $LJ$ potential parameters of the CG system. This ML-based optimization aims to achieve the desired temperature-dependent density, which is consistent with the ground truth (i.e. supplemented experimental and simulation data). 
	The average prediction time for the trained DNN to predict the density at one temperature state from a $2.3$ GHz core $i7$ processor is approximately $0.01$ seconds, whereas the PCG-MD simulation takes an average of $60$ seconds. This pinpoints the speed increment by a factor of $6.2 \times 10^3$ compared to MD simulations, and therefore, significantly accelerates the optimization process. 
	
	From the predicted nonbonded interaction parameters, the resulting local structure i.e. radial distribution function (RDF) $g(r)$ is shown in Figure \ref{Figure:3} (c) along with RDF from the united atom model and $LJ$ potential (dashed line). The radial distribution functions g(r) of the UA-PTMO and CG-PTMO are calculated at $453$ K. The initial peak for both the UA and CG systems is found to correspond to the same radial distance; however, the CG peak value is relatively higher in comparison with the UA system. This is the result of an increased well-depth as predicted by the DNN in the top-down optimization of the nonbonded parameters from different thermodynamic states. The observation of relatively larger peak value in radial distribution function during top-down optimization of nonbonded parameters is consistent with the literature, where top-down approach is used in classical MD for coarse-grained system of polystyrene \cite{hsu2015thermomechanically}. 
	
	\section{CG Model Predictions and Discussion}	
	The performance of DNN accelerated CG potentials is examined by investigating universality and temperature-transferability in two ways. The first way is to compute properties, which have not been included in the training. Transferability indicates the application of the potential parameters for temperatures and molecular weights, which are outside the range of the training set. The machine learned coarse-grained systems (where $N_b$ is the number of beads/chains and $N_c$ is the number of chains) studied in this work are listed in Table \ref{Table:3} along with the ground truth data.
	
	\begin{table}[h!]
		\caption{\label{Table:3} Machine learned coarse-grained (MLCG) PTMO systems studied in this work at $T = 453$ K.}
		\begin{ruledtabular}
			\begin{tabular}{ccccc}
				Model&{$M_n$ (kDa)}&{$N_b$ (beads/chain)}&Chains ({$N_c$)}&{$\rho$ ($g/cm^3$)} \\\hline
				MLCG&$1.8$&$25$&$1000$&$0.868$ \\
				MLCG&$3.6$&$50$&$500$&$0.873$ \\
				MLCG&$7.2$&$100$&$250$&$0.875$ \\
				MLCG&$14.4$&$200$&$125$&$0.876$ \\
				MLCG&$25.2$&$350$&$72$&$0.877$ \\
				MLCG&$36$&$500$&$100$&$0.878$ \\
				MLCG&$72$&$1000$&$100$&$0.879$ \\
				UA&$3.6$&$50$&$50$&$0.875$ \\
			\end{tabular}
		\end{ruledtabular}	
	\end{table}

\begin{figure}
		\centering
		\includegraphics[width=0.5\textwidth]{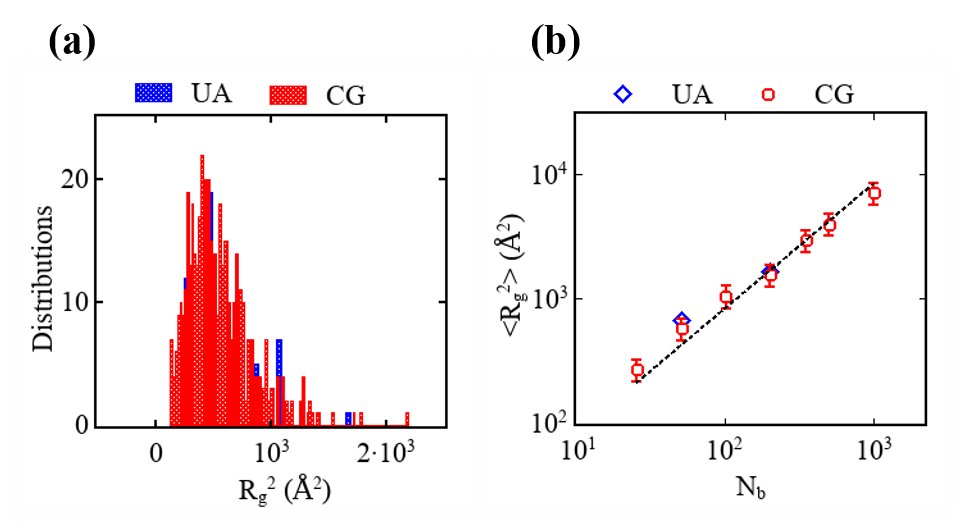}
		\caption{(a) Distribution of squared radius of gyration $R_g^{2}$ is shown for MLCG and UA system for the chain size $N_b = N_m = 50$. (b) The average-mean-square radius of gyration $\langle R_g^{2}\rangle$ as a function of chain length $N_b$ ($T= 453$ K).The diamond symbols represent data from united-atom systems. The dashed line indicates a fit of the power law, $\langle R_g^{2}\rangle \propto {N_b}^{2\nu}$ with $\nu = 1/2$ for ideal chains. }
		\label{Figure:4}
	\end{figure}

	We first study the structural and conformational properties of the machine learned coarse-grained model of PTMO melts through MD simulations. The distribution of squared radius of gyration $R_g^{2}$ of CG and UA models are compared in Figure \ref{Figure:4} (a) for $N_b = 50$. It is to underline that chain size with $50$ beads was included in the training data for potential parameterization. The gyration distributions from the two models are in excellent agreement. We also analyzed the chain dimensions of the coarse-grained PTMO for all the systems tabulated in Table \ref{Table:3}. Figure \ref{Figure:4} (b) visualizes the averaged-mean-squared radius of gyration ($\langle R_g^{2}\rangle$), as a function of chain length $N_b$. Note that $N_b = N_m$ with $N_m$ being the number of monomers (repeat units) in UA system. It can be seen that with increasing chain length of PTMO in the melt, $\langle R_g^{2}\rangle$ is approaching random coil hypothesis predictions (linear $N_b$ dependence, shown by dashed line). Figure \ref{Figure:4} (b) represents the chain statistics well above and below the chains size ($N_b = N_m = 50$), which was included in training and testing. The $\langle R_g^{2}\rangle$ for $N_b = 200$ is in good agreement with chain size of the united atom model at $N_m = 200$, which indicates that the MLCG potential captures the structural features very well. 
	
\begin{figure}
		\centering
		\includegraphics[width=0.5\textwidth]{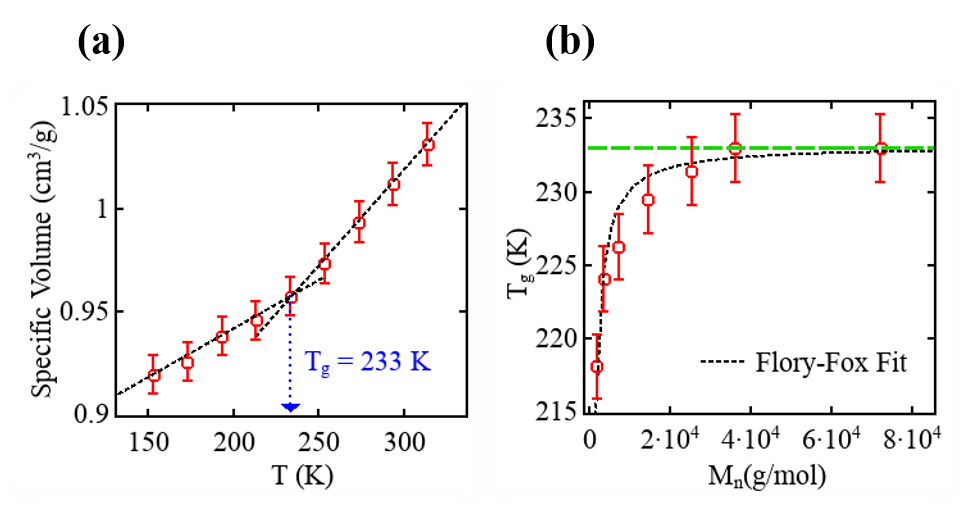}
		\caption{(a) The glass transition temperature is estimated from the specific volume at $N_{b} = 500 $. (b) The limiting behavior of the PTMO is shown by fitting the Flory-Fox equation to the predicted $T_g$ from the developed MLCG potential.}
		\label{Figure:5}
	\end{figure}

	Figure \ref{Figure:5} represents the temperature dependent volumetric behavior along with predicted limiting behavior in terms of glass transition temperature ($T_g$) of the PTMO system. It is important to note that volumetric behavior shown in Figure \ref{Figure:5} (a) is from the temperature range $313$ to $153$ K, which were not included in the nonbonded potential parameterization. The dotted arrow in Figure \ref{Figure:5} (a) indicates the predicted $T_{g} = 233$ K for chain length $N_{b} = 500$. Figure \ref{Figure:5} (b) visualizes the limiting behavior of PTMO melt predicted by MLCG potentials. The predicted glass transition temperatures lie within the range of experimental estimates as tabulated in Table \ref{Table:4}. The limiting behavior is identified by fitting the Flory-Fox relationship to the predicted $T_g$ for the range of molecular weight. The dashed line is to show the experimental observation \cite{rajendran1989synthesis} and the dotted line represents the Flory-Fox fit given by the equation:
	
	\begin{equation}
	{T_g(M_n)} = {T_{g\infty}}({1- {K/{M_{n}}}})
	\label{eq10}
	\end{equation}	
	
	where $T_{g\infty}$ is the glass transition temperature for PTMO chains approaching infinite length and $K$ is an empirical constant and is equal to $130.0$ g/mol from Figure \ref{Figure:5} (b). We consider the model to be "thermodynamically consistent" if it accurately predicts $T_g$ at higher molecular weights. Therefore, the results shown in Figure \ref{Figure:5} are an indication of the transferability of the developed potentials, using our hybrid approach for different thermodynamic states.
	
	\begin{table} 	
		\caption{\label{Table:4} Glass transition temperature predicted by MLCG systems at different molecular weight .}
		\begin{ruledtabular}
			\begin{tabular}{ccc}
				$M_n$ ($g/mol$) &$T_g$ (K) & References \\ \hline
				$1800-72000$& $218.14 -233$& this work \\ 
				$3500-10200$	&$198 - 233$ & \cite{rajendran1989synthesis}	\\
				$44000-44100$	&$198 - 207$ & \cite{ali1993physical}	\\
			\end{tabular}
		\end{ruledtabular}
	\end{table} 	
	
	\begin{figure}
		\centering
		\includegraphics[width=0.25\textwidth]{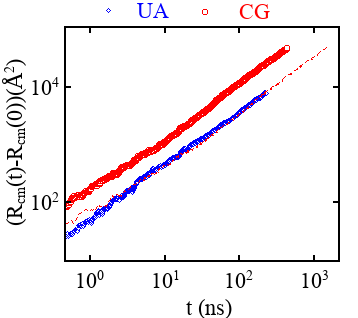}
		\caption{The mean squared displacement (MSD) of the center of mass of CG-PTMO and UA chains are computed at $T = 453$ K. The dashed line represents the time-scaled MSD of CG system with scaling factor $S_{AA‑CG} = 3.35$.}
		\label{Figure:6}
	\end{figure}
	
	To account for the faster dynamics in CG systems Fritz et al. proposed a dynamic scaling factor, $S_{AA‑CG}$ to obtain the quantitative agreement between dynamics in atomistic and CG simulations for one component system. The time scaling factor is the ratio of diffusion coefficients of CG and atomistic system such that $S_{AA‑CG} = D_{CG}/D_{AA}$ and can also be estimated using mean square displacement of beads \cite{fritz2011multiscale}. We estimate the dynamic scaling factor $S_{UA-CG}$ from the linear part of the mean squared displacement of the center of mass ($\langle(R_{cm}(t)-R_{cm}(0))^2\rangle$) as a function of time using the Einstein equation:
	
	\begin{equation}
	{D} = {\lim_{t\to\infty}}\frac{1}{6t}\langle(R_{cm}(t)-R_{cm}(0))^2\rangle
	\label{eq11}
	\end{equation}	
	
	The mean squared displacement of CG and UA system is compared and the dynamic scaling factor ($S_{UA-CG}$) is computed for $N_b = 50$ at $T = 453$ K (chain length and temperature included in training set). In Figure \ref{Figure:6} the mean squared displacement from CG PTMO system, MSD($t_{CG}$) is shifted along the time axis by a feasible time scaling factor $S_{UA-CG} = 3.35$ resulting in an agreement with united-atom data, such that MSD($t_{CG}S_{UA-CG}$) = MSD($t_{UA}$). 
	
	\begin{figure}
		\centering
		\includegraphics[width=0.5\textwidth]{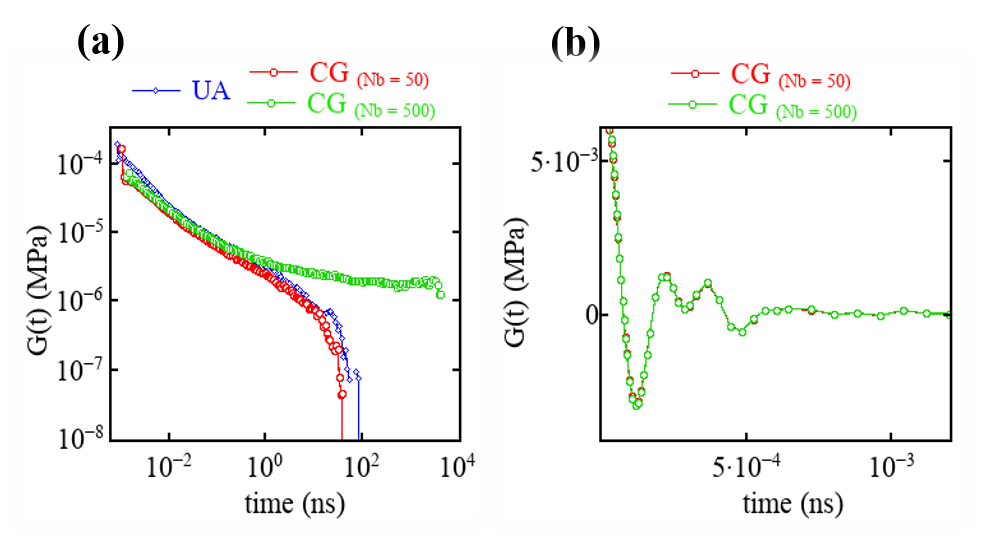}
		\caption{The relaxation spectrum of CG-PTMO and UA chains are computed at $T = 453$ K. (b) The short timescale relaxation behavior PTMO chain lengths below and above entanglement limit.}
		\label{Figure:7}
	\end{figure}
	
	Figure \ref{Figure:7} shows the relaxation spectrum $G(t)$ for machine learned coarse grained system of PTMO along with united atom simulation.  Figures \ref{Figure:7} (a) and (b) shows the long and short time behavior of the CG system, respectively. The stress relaxation spectrum is also computed for chain lengths $N_b = 50$ and $500$ to demonstrate the predictive power of the CG potentials beyond the entanglement length ((experimentally, effective entanglement in polyether diol $\approx 10500$ g/mol corresponding to $145$ PTMO repeating units or $181$ propylene oxide repeating units) \cite{pohl2016dynamics}). As shown in Figure \ref{Figure:6} (a) the agreement between the united-atom model and CG PTMO system is excellent from intermediate time scales to long time scales ($10$ ps $\le$ t $\le$ $10^2$ ns) for the chain length of $50$ beads. 
	
	Generally, for all-atom models sufficient sampling becomes increasingly difficult beyond $1$ ns and stress decays rapidly with higher deviations in relaxation time predictions. However, in this study, the multi-tau correlator method \cite{ramirez2010efficient} is employed to calculate the stress auto-correlation function (SACF) in a canonical ensemble (NVT). The simulation time is increased up to $10^2-10^3$ ns, which provides sufficiently accurate estimates of the relaxation spectrum of the UA model. Importantly, the efficiency of the CG model allows for sufficient sampling well beyond $10^3$ ns, far from the achievable range of atomistic MD simulations. The relaxation spectrum of chain length $N_b = 500$ is an indication of this with appearance of plateau in $G(t)$ up to $10^4$ ns. The comparison between two molecular weights also underlines the consistency of the developed potentials in terms of good agreement from short time scale (Figure \ref{Figure:7} (b)) to intermediate time scale (Figure \ref{Figure:7} (a)). The extended plateau for $N_b = 500$ is due to entanglement. 
	
	\section{Conclusion}
	We proposed a novel framework for development of an efficient, accurate, and transferable coarse-grained (CG) potentials for a model polymer by integrating bottom-up and top-down approaches through machine learning and optimization algorithms. We have demonstrated the versatility of the integrated machine learning method for coarse-graining polyether polymers such as poly(tetramethylene oxide). Using a simple one-bead mapping mechanism, and introducing a deep neural network trained bottom-up optimization to match the atomistic target bonded distributions, we were able to reproduce the structural features of the polymer chain system. In addition, developing a top-down optimization approach for the nonbonded LJ potential parameters allowed us to match the local structure in terms of radial distribution function. We also demonstrated that optimizing the nonbonded parameters by matching the specific volume at multiple temperature points in liquid state provides transferability of potentials to predict the accurate glass transition temperature $T_g$ consistent with experiments. The molecular weight dependent $T_g$ allows the fitting of Flory-Fox equation to predict the limiting behavior of PTMO, which is found to be in good agreement with experimental measurements. This indicates that the developed framework may become a versatile approach to enable transferability of certain thermodynamic state dependent properties in CG systems. Calculated dynamic scaling factor through comparison of the CG and united-atom mean squared displacement of the center of mass, indicates the accelerated speed of CG systems. By comparing the actual computer wall-time of the UA and CG simulations ran for $50$ monomer systems, we found out that the performance of UA system is $9.084$ ns/day and for CG system $1735.109$ ns/day on a computer of $28$ CPUs with average CPU speed of $98.9$\%. This means that the CG model of PTMO provides a speed-up of over $200$ times compared to the united-atom model due to the lower degrees of freedom and larger time step. The CG model enables prediction of the long-time behavior, of particular importance for capturing the full stress relaxation behavior of the polymeric systems. This is due to the increased efficiency and faster dynamics of the CG chains. 
	
	Therefore, in the present work, we established the advantages of using a hybrid MLCG method for polymer coarse-graining over the classical coarse-graining methods. Using our novel methods, first, we are able to demonstrate simultaneous and accurate prediction of structural, dynamics and experimentally consistent glass transition along with limiting behavior, which has been elusive to accomplish in traditional coarse-graining methods. Second, our framework enables the transferability of the developed potentials to ensure accurate thermomechanical predictive behavior by including the polymer chain length well beyond the oligomers, in the parameterization (training, testing and optimization) process. Third, by incorporating experimental and atomistic data of temperature-dependent density into nonbonded parameterization, the accuracy is enhanced (i.e., experimentally consistent $T_g$ and limiting behavior are achieved) while ensuring the generalizability for thermodynamic state dependent properties of CG models. Finally, the protocol has been shown to be robust, and having been applied to poly(tetramethylene oxide), it could potentially be applied to other polymeric systems.

	\begin{acknowledgments}
		This research is supported by the Commonwealth of Australia as represented by the Defence Science and Technology Group of the Department of Defence.  We are grateful and acknowledge Prof. Saman Halgamuge for insightful discussions throughout this work. We also acknowledge Spartan-HPC and Argali-HPC at Faculty of Engineering and Information Technology, The University of Melbourne for providing us the computing facility.
	\end{acknowledgments}
	
	\section*{Data Availability Statement}
	The authors confirm that the data supporting the findings of this study are available within the article and the supplementary materials. Supporting materials include LAMMPS input files and starting configuration files for UA and CG polymer structures at all values of temperature and molecular weight, an excel file with the input values and output responses for the target UA simulations and the CG simulations. Resources available at {link}. Additional data are available from the corresponding author Dr Ellie Hajizadeh upon reasonable request.
	
	\section*{Codes Availability Statement}	
	Input files for the open source software LAMMPS and all necessary parameters needed to implement the MLCG model are provided in the repository {link}. Additional details on the code used are available from the corresponding author Dr Ellie Hajizadeh upon reasonable request.
	
	\section*{References}
	\nocite{*}
	\bibliography{aipsamp}
	
	\appendix
	
	\section{MD Simulation Details}  \label{UAS}
	We performed united-atom (UA) molecular dynamics simulations to generate reference (the ground truth) structural distributions of PTMO chains in the melt state for the purpose of parameterizing the potential parameters of the CG bead-chain model. Generally, for polymeric materials, the CG potentials are derived from simulations of shorter chains or oligomers with the assumption that the derived potentials are transferable to longer length chains \cite{agrawal2014simultaneous}. In this study, we use UA chains of PTMO containing $50$ repeating units, equivalent to $250$ atoms and united atoms along the chain for generating the structural distributions.  Considering the computational limitations of the UA-PTMO MD simulations we selected the system of $50$ such chains for collecting data as reference. 
	
	The repeat unit $-(CH_2CH_2OCH_2CH_2)-$ of the UA-PTMO chain is defined in such a way that every oxygen and every methylene group is treated as an atom and united atom, respectively. The UA potential used in the reference system is the Transferable Potentials for Phase Equilibria-UA (TraPPE-UA) for the intra (bonded) and inter-molecular (nonbonded) interactions developed by Lempesis et. al  \cite{lempesis2016atomistic}. Before collecting data the system was equilibrated at $T = 453$ K for $5$ ns in a canonical ensemble using the deterministic Nose-Hoover thermostat with a time constant equal to $0.1*dt$ ps \cite{nose1984unified,hoover1985canonical}.
	
	We measure two types of structural distributions as well as temperature-dependent density of the melt PTMO from our UA simulations. First, we collect the bond length $l$ distribution data, i.e. the distances between "virtual" bead centers along the UA-PTMO chain, representing centers of the adjacent beads in the CG-PTMO model. Given that all the beads in the CG chains are of the same type, there is only one type of chemical bond in CG system, thus, only one set of bond length distribution data is recorded. This distribution is used to parameterize the bond stretching potential (the functional form of the potential is given in Table \ref{Table:2}). Second, we collect bond angle $\theta$ data and its distribution between three consecutive CG beads, and this distribution is used to parameterize the CG-PTMO bending potential (the functional form of the potential is given in Table \ref{Table:2}). 
	
	Finally, we measured the temperature-dependent density of melt PTMO at $P = 1$ atm by quenching the equilibrated UA-PTMO from $453$ K to $313$ K in decrements of $20$ K. The measured temperature-dependent density is used to optimize the nonbonded CG-PTMO potentials to ensure their transferability to a higher molecular weight and/or lower temperature ranges.

\end{document}